\documentclass[aps,prl,twocolumn,showpacs,superscriptaddress,groupedaddress]{revtex4}  
\usepackage{graphicx}  
\usepackage{dcolumn}   
\usepackage{bm}        
\usepackage{amssymb}   
\usepackage{amsmath} 
\usepackage{xcolor} 
\usepackage{caption} 
\usepackage{natbib} 
\begin{document}

\title{Observations of Heteroclinic Bifurcations in Resistive MHD Simulations of the Plasma Response to Resonant Magnetic Perturbations}
\author{T. E. Evans}
\author{W. Wu}
\affiliation{General Atomics, P.O. Box 85608, San Diego, California 92186-5608, USA}
\author{ G. P. Canal}
\affiliation{Department of Applied Physics, University of S\~{a}o Paulo, S\~{a}o Paulo, CEP 05508-090, Brazil}
\author{N. M. Ferraro}
\affiliation{Princeton Plasma Physics Laboratory, PO Box 451, Princeton, New Jersey 08543-0451, USA}

\date{\today}

\begin{abstract}
A new class of static magnetohydrodynamic (MHD) magnetic island bifurcations is identified in rotating spherical tokamak plasmas during single- and two-fluid resistive MHD simulations. As the magnitude of an externally applied non-axisymmetric magnetic field perturbation is increased in these simulations, the internal flux surfaces that make up a sub-set of the resonant helical magnetic islands in the plasma gradually elongate and undergo heteroclinic bifurcations. The bifurcation results in the creation of a new set of hyperbolic-elliptic fixed points as predicted by the Poincar\'e-Birkoff fixed point theorem. Field line calculations without including the resistive MHD plasma response to the applied perturbation field do not undergo this class of bifurcations indicating the importance of plasma self-organization in the bifurcation process. 
\end{abstract}


\maketitle

Resistive magnetohydrodynamic (MHD) theory and simulations are fundamental tools for understanding the physics of topological bifurcations in the trajectories of magnetic field lines that are associated with tearing and reconnection in astrophysical \cite{bran05}, solar \cite{sant07}, space \cite{treu13} and magnetically confined toroidal plasmas \cite{ferr12}. While magnetic tearing and reconnection theory is conceptually well established and widely accepted, there are open questions concerning the detailed dynamics responsible for these topological changes that are not yet fully understood \cite{gonz16}. In resonantly perturbed toroidal magnetic plasmas, such as those in tokamaks, theory predicts that only three parameters are required to determine the topology of a fully reconnected magnetic state, namely the normalized plasma viscosity, rotation, and resistivity \cite{fitz98}. As discussed in this Letter, linear response studies of resonantly perturbed tokamak plasmas with the \textsc{m3d-c$^1$} resistive MHD code \cite{ferr13} have resulted in the discovery of a new class of magnetic field line bifurcations that do not fit into the topological framework of this three-parameter bifurcation theory. These results are particularly important for understanding the underlying physics of self-organization in toroidal plasmas. The topology of the magnetic equilibrium field in toroidal plasmas plays an essential role in determining the properties of the energy, particle and momentum confinement as well as its MHD stability. Therefore, understanding the mechanisms responsible for this new class of topological bifurcations is of particular importance as the plasma parameters in high-power toroidal fusion devices approach those needed to achieve self-sustained burning or ignited states, since the ability of the plasma to self-organize can open access to new types of operating regimes.

Experience has shown that as the magnetic topology changes in tokamaks, unexpected types of plasma dynamics can appear. In particular, bifurcations of the equilibrium magnetic topology are known to be associated with the ability of the plasma to self-organize in unanticipated ways. For example, the spontaneous generation of a bifurcated helical magnetic core equilibria and saturated internal kinks, observed in several conventional aspect ratio tokamaks \cite{well87,pecq97} and in a small aspect ratio spherical tokamak \cite{chap10,mena05}, are found to be associated with peaked pressure profiles that trigger bifurcated MHD equilibrium states reminiscent of saturated internal kink modes found in \textsc{animec} simulations \cite{coop13}. In the RFX-mod reversed field pinch, it is found that as the plasma current is increased, the hyperbolic (x-) point of the dominant core magnetic island merges with the main magnetic axis to form a new self-organized quasi-single-helicity state \cite{lore09}. Core magnetic islands have also been found to trigger internal transport barriers \cite{inag10} and to spontaneously generate cyclical dynamics in the heat transport across low-order rational surfaces near the mid-radius of the DIII-D tokamak and the large helical device (LHD) \cite{kida16}. Consequently, understanding the underlying physics responsible for triggering self-organized plasma states is of paramount importance for the development of magnetic fusion energy devices and is of intrinsic interest from a broader scientific perspective.

In magnetically confined toroidal plasmas, such as in tokamaks and stellarators, non-axisymmetric magnetic field perturbations result from vacuum field sources external to the plasma and from internal MHD plasma instabilities. As $\beta$ increases, changes in the equilibrium magnetic topology, plasma stability and plasma transport become highly sensitive to the properties of the external non-axisymmetric vacuum field perturbations. Here, $\beta$ is the ratio of the volume averaged plasma pressure $\langle\emph{p}\rangle$ to the magnetic pressure $B^2/2\mu_0$ and the associated normalized plasma beta is defined as $\beta_N=\beta(aB_T/I_p)$, where \emph{a} is the minor radius of the plasma, $B_T$ is toroidal magnetic field and $I_P$ is the plasma current. Changes in the resonant and non-resonant magnetic equilibrium topology, caused by vacuum field perturbations, are referred to as the plasma response. These changes are known to alter neoclassical tearing mode (NTM) and edge localized mode (ELM) stability as well as the particle, energy and momentum confinement of the plasma \cite{evan15}. Coupling between stable ideal kink modes and resonant modes on rational surfaces with $m=nq$ have been shown to correlate with reductions in the edge pressure gradient resulting in changes to the ELM stability \cite{pazs16,evan06}. Here, \emph{m} and \emph{n} are poloidal and toroidal mode integers respectively, and q($\psi_N$), a function of the normalized poloidal magnetic flux ($\psi_N$), is the safety factor associated with the radial variation in the normalized poloidal flux $\psi_N$ of the axisymmetric equilibrium magnetic field. Thus, the coupling between stable ideal kink modes, with (\emph{m,n}) components that reside at a smaller radius than the resonant q($\psi_N$) surface, can significantly modify the resonant plasma response to an external magnetic perturbation field.

In this Letter, the plasma response to $n=3$ externally applied non-axisymmetric magnetic perturbation fields ($\delta{b}^{ext}=\delta{b}_r^{ext}+\delta{b}_\theta^{ext}+\delta{b}_\phi^{ext}$) is simulated in two high $\beta_N$ spherical tokamak discharges using the \textsc{m3d-c$^1$} resistive MHD code. These simulations result in the formation of $n=3$ heteroclinic magnetic islands on each low-order rational surfaces from the center of the plasma to the edge. As $\delta{b}_r^{ext}$ is increased in the simulation, it is found that magnetic islands on resonant surfaces where $\emph{m}/3$ is a rational number undergo internal topological bifurcations. In these simulations, the total resonant magnetic field in the plasma is $\delta{b}^{total}=\delta{b}^{ext}+\delta{b}^{plasma}$. Here, $\delta{b}^{plasma}$ is the MHD plasma response field due to the application of $\delta{b}^{ext}$. It is found that the threshold value of  $\delta{b}^{ext}$ required for $\emph{m}/3$ islands to undergo a bifurcation depends on $\beta_N$. Furthermore, in cases where no plasma response is included, i.e., $\delta{b}^{plasma}=0$ no bifurcations are observed.

The linear time-independent single- and two-fluid \textsc{m3d-c$^1$} simulations in this work were carried out using synthetically generated kinetic Grad-Shafranov \cite{llao05} plasma equlibria in the NSTX-U spherical tokamak \cite{mena17}. Two equlibria are simulated, one with $\beta_N=5.5$, $I_p=1.5$ MA, $B_T=1.0$ T and a q($\psi_N$) on the 95\% $\psi_N$ surface of $q_{95}=8.7$ and another with $\beta_N=7.6$, $I_p=2.0$, $B_T= 1.0$ and $q_{95}=5.5$, where q($\psi_N)>1$ everywhere in both cases.  Non-axisymmetric vacuum field perturbations are applied using 12 equally spaced toroidal rectangular coils above and below the equatorial plane, referred to as the Non-axisymmetric Control Coil (NCC). This coil set is being considered for instillation in NSTX-U over the next few years. In the simulation discussed here, the normalized currents in this coil have a toroidal distribution pattern in the upper NCC of (+1, 0, -1, 0, +1, -1, 0, +1, 0, -1, 0, 0) with opposite signs in the lower NCC. \textsc{m3d-c$^1$} simulations are done with currents in the NCC ($I_{NCC}$) ranging from 1 kA to 30 kA using the radial plasma profiles shown in figure 1 as input.

\begin{figure}[h]
\includegraphics[scale=0.11]{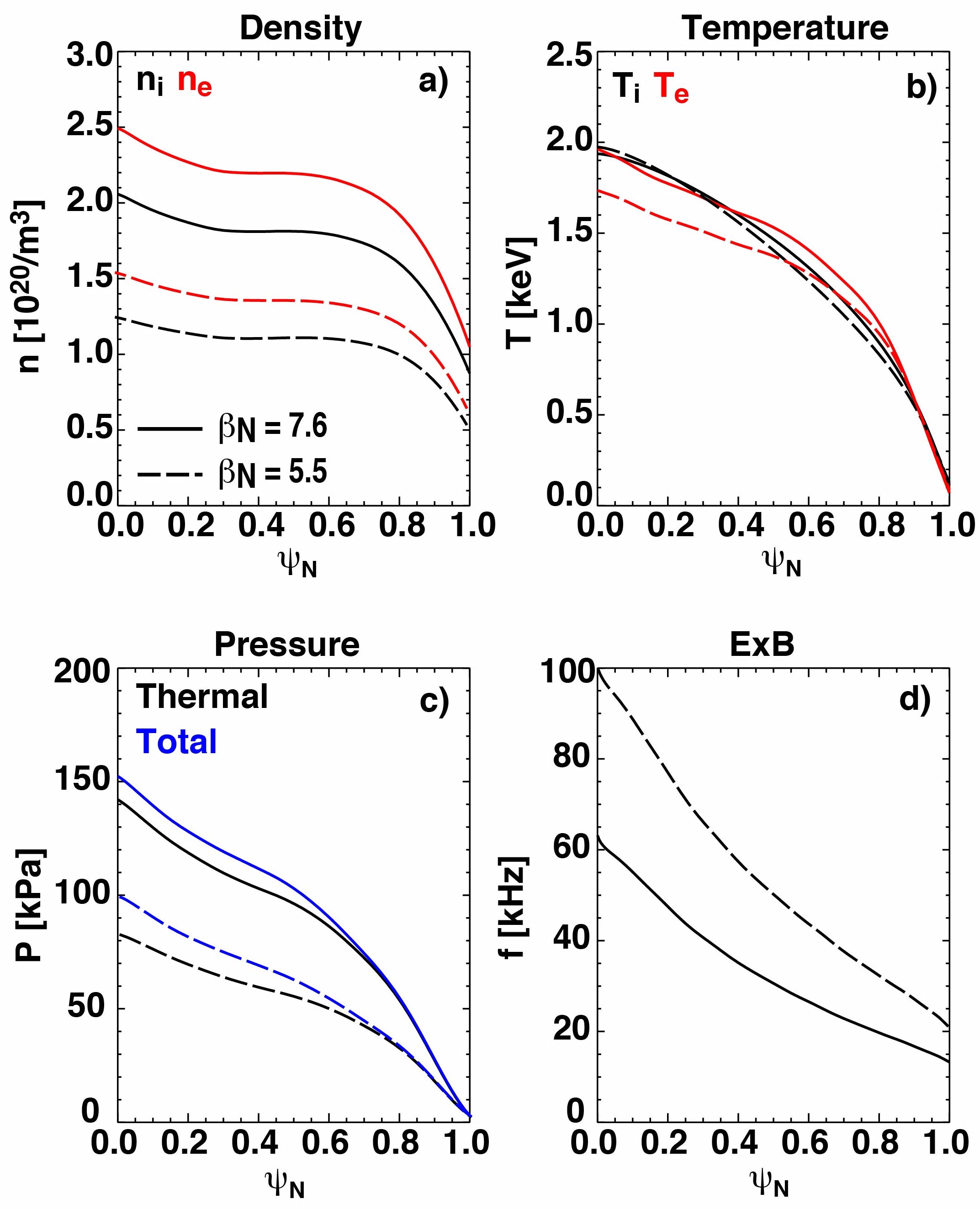}
\caption{a) ion (black) and electron (red) density, b) ion (black) and electron (red) temperature, c) thermal (black) and total (blue) pressure, and d) poloidal \emph{E}$\mathsf{x}$\emph{B} rotation ($\omega_{ExB}=2\pi{f_{ExB}}$) for $\beta_N=7.6$ (solid line) and $\beta_N=5.5$ (dashed line) cases.}
\end{figure}

Due to the relatively high poloidal \emph{E}$\mathsf{x}$\emph{B} rotation in these discharges, shown in figure 1(d), the \textsc{m3d-c$^1$} simulations show that $\delta{b}^{plasma}$ is reduced by at least a factor of 2 on the resonant surfaces of interest compared to that of $\delta{b}^{ext}$. This results in small magnetic islands, with widths $\Delta\psi_N^{island}<0.02$, formed on each rational surface corresponding to the fundamental $n=3$ toroidal modes produced by the NCC.

\begin{figure*}
\includegraphics[scale=0.55]{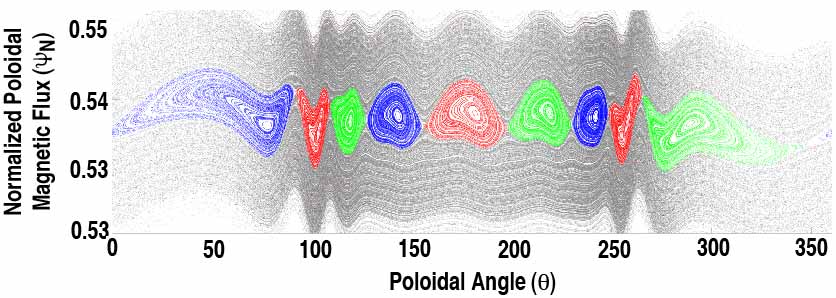}
\caption{Three isolated sets of $m/n=$ 3/1 heteroclinic magnetic islands located on the q=3 normalized poloidal flux surface in NSTX-U during the application of a 2 kA $n=3$ NCC perturbation field.}
\end{figure*}

In the unperturbed $\delta{b}^{ext}=0$ simulations, field lines are confined to the surface of a 3D torus defined by $\psi_N$, the normalized toroidal flux ($\chi_N$), the poloidal angle ($\theta$) and the toroidal angle ($\phi$), where $\psi_N$,$\theta$,$\chi_N$,$\phi$ are referred to as action-angle variables. Using this representation, the action $\psi_N$, when plotted as a function of the poloidal angle $\theta$, describes a set of parallel straight field lines that intersect a 2D Poincar\'e plane at a $\phi$ = constant angle. On rational surfaces with $\delta{b}^{ext}\neq$ 0, field lines form periodic trajectories which result in a finite number of fixed-points when projected on the Poincar\'e plane. For example, on the $q(\psi_N)=3$ surface individual field lines undergo 3 toroidal revolutions ($\phi=6\pi$) before returning to their original $\psi_N$,$\theta$ position. Here, each fixed-point undergoes 3 toroidal transits in order to complete 1 poloidal transit resulting in a winding number of $m/n=$ 3/1. Since there are 3 set of these isolated islands on $q(\psi_N )=3$, this results in an effective $m/n$ winding number of 9/3. The application of a small $\delta{b}^{ext}$ field destroys the toroidal symmetry resulting in the formation of an even multiple of fixed-points as prescribed by the Poincar\'e-Birkhoff fixed-point theorem \cite{birk27}.

According to the Poincar\'e-Birkhoff fixed-point theorem, an even number of fixed-points must appear on every rational surface for an arbitrarily small $\delta{b}^{ext}$. In conservative area or magnetic flux preserving ($\nabla\cdot\emph{B}=\nabla\cdot\nabla\mathsf{x}\emph{A}=0$) Hamiltonian systems, where each rational surface is bounded on both sided by an irrational surface, the KAM theorem \cite{mose73}, in conjunction with the Poincar\'e-Birkhoff fixed-point theorem mandates that each pair of fixed-points must consist of a both a hyperbolic and an elliptic fixed-points. In general, this allows for the formation of $N_{fp}=2\ell$\emph{m} fixed-points, where \emph{m} is the period of each fixed-point, $\ell$ is referred to as the topological index corresponding to the number of isolated fixed-point pairs and 2 is the Poincar\'e-Birkhoff integer. Since topological island bifurcations must preserve the fixed-point winding number on each resonant surface, $\ell=1$ for homoclinic islands and $\ell>1$ for heteroclinic islands. An example of a set of heteroclinic islands resulting from the \textsc{m3d-c$^1$} $\beta_N=5.5$ simulation on the $q(\psi_N)=3$ surface in the NSTX-U with $I_{NCC}=2.0$ kA is shown in figure 2.

This figure is obtained by integrating the nonlinear magnetic field line differential equations produced by $\delta{b}^{total}$ using the TRIP3D code \cite{evan02}. Here, 3 isolated heteroclinic sets ($\ell = 3$) of period 3 ($m = 3$) islands are formed when the NCC is turned on, each of which is comprised of 3 elliptic fixed-points (o-points) and 3 hyperbolic fixed-points (x-points), resulting in $N_{fp}=2\ell\emph{m}=18$. Field lines originating on each fixed-point make $m = 3$ toroidal revolution before returning to their original $\psi_N,\theta$ position. As shown by the red, blue and green color coding in figure 2, field line trajectories corresponding to one set of heteroclinic islands do not connect to either of the other two heteroclinic island sets.

\begin{figure*}
\includegraphics[scale=0.15]{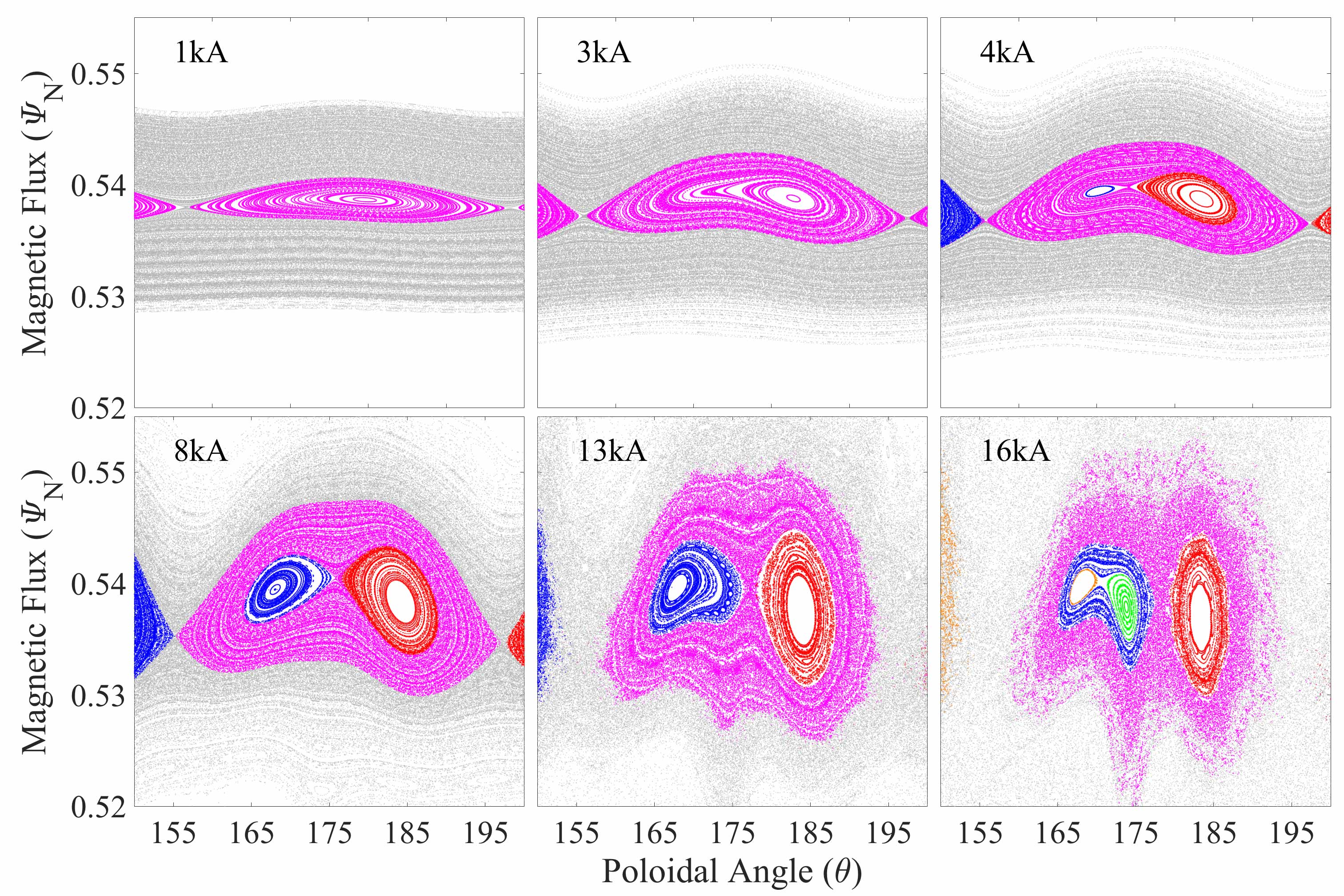}
\caption{Topological change in the inner flux surfaces $q(\psi_N) = 3$ magnetic islands $I_{NCC}$ is increased from 1 kA to 16kA.}
\end{figure*}

Figure 3 shows a detailed view of the $q(\psi_N) = 3$ MHD magnetic island with its elliptic fixed-point at $\theta = 179.9^{\circ}$ in figure 2 for increasing $I_{NCC}$ levels. The increase in $I_{NCC}$ causes a dramatic change in the topological structure of the inner flux surfaces of the island along with all of its heteroclinic partners and neighbors. Initially, $\Delta\psi_N^{island}$ increases with $I_{NCC}$ and the elliptic fixed-point moves from $\theta = 179.9^{\circ}$ to $\theta = 183.1^{\circ}$ while a sub-set of internal island flux surfaces stretch asymmetrically in the opposite direction toward the screened vacuum island elliptic point located at $\theta = 163.3^{\circ}$, as shown in the $I_{NCC} = 3.0$ kA panel of figure 3. This asymmetric internal flux surface stretching is consistent with a change in $\delta{b}^{plasma}$ when linearly superimposed on the increasing $\delta{b}^{ext}$ field, and is expected to involve a redistribution of the $\delta{b}^{ext}$ screening current.  It is noted that the position of MHD island hyperbolic fixed-points remains constant during this process. As shown in the $I_{NCC} = 4.0$ kA panel, the elongated lobe has bifurcated to form a new set of MHD island elliptic and hyperbolic fixed-points located at $\theta = 170.0^{\circ}$ and $\theta = 173.9^{\circ}$ respectively. We refer to this process as a \emph{heteroclinic MHD magnetic island bifurcation} since it results in the formation of 3 new heteroclinic sets of isolated $m = 3$ islands, with $\ell_b = 2\ell = 6$ resulting in $N_{fp} = 2\ell_bm = 36$, as required to preserve the $m/n=$ 3/1 winding number in each set of islands on this resonant surface.  The bifurcation occurs at $I_{NCC} = 3.48$ kA.

As $I_{NCC}$ continues to increase beyond 3.48 kA the original MHD island elliptic fixed-point moves to $\theta = 184.5^{\circ}$ while the new elliptic fixed-point moves to $\theta = 168.2^{\circ}$ and the new hyperbolic fixed-point moves to $\theta = 175.7^{\circ}$. These fixed-points continue to move poloidally by small amounts as $I_{NCC}$ is increased to 14.7 kA where a second heteroclinic bifurcation takes place. This second bifurcation results in 3 new heteroclinic sets of isolated m=3 islands, with $\ell_{2b} =1.5\ell_b = 9$ resulting in $N_{fp} = 2\ell_{2b}m=54$. These heteroclinic bifurcations also occurs on each of the period 3 rational surfaces studied in this simulation starting at $q(\psi_N)=5/3$ and cascading radially outward, down the pressure profile shown in figure 1c, as $I_{NCC}$ is increased. A similar process is observed during simulations of the $\beta_N = 7.6$ case but in this case only one bifurcation occurs on the  $q(\psi_N) = 3$ surface when $I_{NCC}$ passes through 4.23 kA as it is increased to 30 kA.

Although not visible in figure 3 due to the generation of a screening current by the MHD plasma response, the vacuum heteroclinic island due to $\delta{b}^{ext}$ is located inside the MHD island with its elliptic and hyperbolic fixed points on the $q(\psi_N) = 3$ surface at $\theta = 163.3^{\circ}$ and $\theta = 183.5^{\circ}$ respectively. Here, the hyperbolic fixed-point of the vacuum island is located at approximately the same position as the original elliptic point of the MHD island shown in figure 3 with $I_{NCC}= 1$ kA. Our hypothesis is that the position of the vacuum island fixed-points along with the non-resonant kink aligned perturbation field influences the bifurcation dynamics of the MHD islands, as the NCC current is increased, by: 1) regulating the screening current needed to prevent the vacuum island from opening and 2) distorting/stretching the flux surfaces. Coupling between the non-resonant kink response, in the vicinity of each rational surface, and the resonant island response is believed to be the mechanism responsible for these bifurcations.

The shift between the vacuum and MHD island elliptic fixed-points is expected to result from a redistribution of the plasma current density inside the MHD islands. This is required to maintain the co-$I_{p}$ vacuum island screening current density filament ($j_{\parallel-vac}$) while generating a counter-$I_{p}$ current density filament ($-j_{\parallel-MHD}$)  needed to open the MHD island elliptic points. As $I_{NCC}$ increases, $j_{\parallel-vac}$ must increase in order to maintain a screening current consistent with $\omega_{\emph{E}\mathsf{x}\emph{B}}$. A secondary, counter-$I_{p}$, $-j_{\parallel-MHD}$ filament located in the elongated flux surfaces between the screened vacuum and original MHD island elliptic fixed-points is formed as $I_{NCC}$ increases. Eventually, $\delta{b}^{plasma}$ due to the secondary $-j_{\parallel-MHD}$ filament cancels $\delta{b}^{plasma}$ from the original $-j_{\parallel-MHD}$ filament at a point on the $q(\psi)=3$ surface between the two $-j_{\parallel-MHD}$ filaments and a new set of hyperbolic fixed-points with their associated Poincar\'e-Birkhoff elliptic fixed-points are formed. It is also found that linear two-fluid \textsc{m3d-c$^1$} simulations produce qualitatively equivalent results to those of the single-fluid simulations indicating that the Hall term in the \textsc{m3d-c$^1$} Ohm\textsc{\char13}s law equation does not affect the bifurcation process.

In addition to satisfying the Poincar\'e-Birkhoff fixed-point theorem, linear \textsc{m3d-c$^1$} simulations demonstrate that the dynamics of these heteroclinic island bifurcations are the manifestation of a smooth/continuous, invertible, process controlled by $\delta{b}^{ext}$ that does not result in bistable or hysteresis-like behavior associated with various types of tearing modes \cite{cass10}. This behavior is consistent with degenerate equilibrium points associated with local saddle-node bifurcations that are found in Hamiltonian systems composed of conservative vector fields \cite{guck02}. In addition, the process involved in the stretching of the internal MHD island flux surfaces, leading up to the these heteroclinic bifurcations, is consistent with a diffeomorphic type of evolution found in a conservative dynamical system where subsets of differentiable manifolds undergo smooth one-to-one, invertible, topological transformations as a control parameter is varied. 

In summary, a new class of heteroclinic magnetic island bifurcations has been identified during variations of an externally applied non-axisymmetric perturbation field in linear single- and two-fluid \textsc{m3d-c$^1$} resistive MHD simulations of two spherical tokamak plasmas. These bifurcations preserve the rational flux surface winding numbers by generating new sets of isolated (heteroclinic) magnetic islands inside existing heteroclinic MHD magnetic islands starting on rational surfaces near the center of the plasma and cascading outward as the perturbation field $\delta{b}^{ext}$ is increased. Simulations with different normalized plasma pressures ($\beta_N$) result in similar bifurcation sequences but with different values of the $\delta{b}^{ext}$ bifurcation parameter. As $\beta_N$ is increased, bifurcations occur at larger $\delta{b}^{ext}$ values. Based on experimental results during the presence of magnetic islands in tokamaks and stellarators, bifurcations of this type can be expected to result in significant plasma transport and stability changes as well as possible observations of new types of self-organized behaviors.

\section*{Acknowledgements}

This work was supported by the US Department of Energy under DE-SC0012706, DE-SC0018030 and DE-AC02- 09CH11466.

\end{document}